\documentclass[preprint,showpacs,preprintnumbers,amsmath,amssymb]{revtex4}
\usepackage{graphicx}
\usepackage{dcolumn}
\usepackage{bm}
\begin{document}
\title{Exploring metastability via the third harmonic measurements in single crystals of $2H$-$NbSe_2$ showing anomalous Peak effect}

\author{A. D. Thakur$^{1,~\star}$, S. S. Banerjee$^2$, M. J. Higgins$^3$, S. Ramakrishnan$^1$, A. K. Grover$^{1,~\star}$ }
\affiliation{$^1$ Department of Condensed Matter Physics and Materials Science, Tata Institute of Fundamental Research, Mumbai 400005, India \\
$^2$ Department of Physics, India Institute of Technology-Kanpur, Kanpur 208076, India \\
$^3$ NEC Research Institute, Princeton, New Jersey 08540, USA}
\date{\today}

\begin{abstract}
We explore the metastability effects across the order-disorder transition pertaining to the peak effect phenomenonon in critical current density ($J_c$) via the first and the third harmonic ac susceptibility measurements in the weakly pinned single crystals of $2H$-$NbSe_2$. An analysis of our data suggests that an imprint of the limiting (spinodal) temperature above which $J_c$ is path independent can be conveniently located in the third harmonic data ($\chi_{3\omega}^{\prime}$).
\end{abstract}
\maketitle


The ubiquitous occurence of an anomalous enhancement in critical current density ($J_c(H, T)$) in weakly pinned superconducting samples with the increase in applied field (H) or temperature (T) is termed as the peak effect (PE) \cite{1,2}. There is a widespread belief that the PE represents a dynamical transition (e.g., from elastic flow to plastic flow) in the driven vortex matter \cite{3, 4, 5}. Using STM imaging measurements on a very weakly pinned single crystal of the well studied $2H$-$NbSe_2$ system, Troyanovsky {\it et al} \cite{6} have shown that the collective motion of neighboring vortices gives way to positional fluctuations of individual vortices just above the onset temperature ($T_p^{on}$) of the PE.

As regards the nature of the phase transition across the PE region, X. S. Ling {\it et al} \cite{7} have sought to provide direct structural evidence in favor of the first order nature of the order-disorder transition commencing at $T_p^{on}$ in a crystal of Nb. An independent evidence in support of the first order nature of the phase transition in typically weakly pinned single crystals of $2H$-$NbSe_2$ came via bulk transport studies \cite{8, 9} and was confirmed via local scanning ac micro-Hall bar microscopy \cite{10} from the visualization of an interface separating the weaker pinned (ordered) and stronger pinned (disordered) regions across the PE. The first order phase transition notion appears further fortified with the recent results of Xiao {\it et al} \cite{11}, who explored different (partially ordered) metastable states above the usually ascertained $T_p^{on}$. Their transport  measurements essentially amount to locating higher peak temperatures ($T_p$) for superheated metastable states. The highest (limiting) value of the $T_p^{max}$ for a given field can be seen to lie close to their spinodal temperature, $T_s$; above it, the $J_c$ is ascertained to be single valued (i.e., no history effects) and it monotonically decreases with further increase in temperature.

An objective of our present work is to make a proposition for the demarcation of the spinodal line for the order-disorder transition in the underlying (static) vortex matter using the contact-less ac susceptibility measurements. One advantage of the ac susceptibility measurements is that the vortex array is shaken by a tiny driving force, whereas in the usual transport experiments, to observe the PE, the vortex array has to be driven with forces much higher than the critical pinning force.

We have focussed our attention on the single crystals of $2H$-$NbSe_2$. Earlier ac susceptibility studies in them \cite{12, 13, 14} have elucidated the correlation between the quenched random pins, effective disorder and the metastability/history effects. The PE anomaly can be observed to be sharper than the superconducting transition width in a {\it nascent} pinned $2H$-$NbSe_2$ crystal, and {\it history effects in such a sample could be easily swamped with a tiny ac driving force} \cite{15}. However, for typically weakly pinned samples, the PE phenomenon prior to the $T_c(H)$ comprises two distinct fracturing steps, across both of which the metastability effects develop prominently \cite{16}. It is further noted \cite{15} and recalled here that with the progressive enhancement in quenched disorder, there appears an additional anomalous peak in $J_c(H, T)$ response prior to the classical peak effect located at the edge of the $T_c(H)$ line. The additional peak imprints as the second magnetization peak (SMP) in the isothermal magnetization hysteresis (M-H) loops \cite{17}. It suffices to state here that the two anomalies appear distinct and different. The circumstances where they could manifest close to each other and/or overlap, and the dynamical behavior across them could get admixed, shall be dealt with  elsewhere \cite{18}.

The ac susceptibility experiments have been performed using a home built \cite{19} ac susceptibility setup, in the frequency interval 21 Hz to 211 Hz, and for ac field ($h_{ac}$) amplitude lying in the range of 0.5~Oe to 2.5~Oe (rms). The $2H$-$NbSe_2$ samples are the crystals $Y^{\prime}$ ($T_c(0) \approx 7.25 K$) and $Z$ ($T_c(0) \approx 6 K$), studied earlier by S. S. Banerjee {\it et al} \cite{12, 13, 14, 15, 16}, alongwith another crystal $Z^{\prime}$ ($T_c(0) \approx 6 K$). The isofield measurements have been performed in three modes: zero field cooled (ZFC) warm-up, field-cooled cool down (FCC) and the field-cooled warm-up (FCW).

Fig.1 shows the in-phase ac susceptibility ($\chi_{\omega}^{\prime} (T)$) data in different modes in the crystal $Y^{\prime}$  in a field of 15 kOe. The imprint of the PE in it is self evident: $\chi^{\prime} \sim -1 + \alpha h_{ac} / J_c(H, T)$, where $\alpha$ is geometry and size dependent factor \cite{20}. A pertinent thing to note in these data is the distinction between the two field cooled warm-up responses. The two initial FC states (FC1 and FC2) were obtained while cooling down in a dc field with superimposed ac field $h_{ac}$ kept {\it on} and {\it off}, respectively. When $h_{ac}$ of 0.5~Oe (rms) remained present, there appeared little difference between the initial field cooled (FC1) state and the ZFC state, and the subsequent warm-up responses of both of them. However, when the $h_{ac}$ remained switched off, there is a discernible difference in the initial FC2 and the ZFC state. The difference during subsequent warm-up appears to survive upto (or just above) the peak temperature $T_p$ of the PE. These data atest to the correlation between the pinning and the exploration of the metastable states prior to the PE \cite{12}, and the role of an ac driving force in annealing the disordered state by the shaking effect \cite{21}.

Figs. 2(a) to 2(c) show the $\chi_{\omega}^{\prime} (T)$ data in a somewhat stronger pinned crystal $Z^{\prime}$ (i.e., where the pinning effects are larger than the crystal $Y^{\prime}$) and in which the history effects manifest in a robust manner such that keeping $h_{ac}$ on or off during cool down does not affect the FC state. Note first the presence of two anomalous variations in $\chi_{\omega}^{\prime} (T)$, which we believe represent the SMP and the PE anomalies. In the $\chi_{\omega}^{\prime} (T)$ data at 2.6~kOe in the ZFC mode, one can also identify the two characteristic steps near $T_p^{on}$ and $T_p$ values, as reported earlier by Banerjee {\it et al} \cite{16}. At higher fields, the stepwise fracturing feature across the PE gradually transforms into a continuous amorphization process from $T_p^{on}$ to the peak temperature $T_p$. An interesting behavior in the data at 6~kOe and 10.5~kOe (cf. Figs. 2(b) and 2(c)) is that the peak temperature of the PE is different in different modes, the highest value being in the ZFC mode, $T_p^{zfc}$. To emphasize, the $\chi_{\omega}^{\prime} (T)$ responses in Figs. 2(b) and 2(c) appear to mimic the scenario that emerges from the critical current $I_c(T)$ data presented for different sample histories in a crystal of $2H$-$NbSe_2$ by Xiao {\it et al} (cf. Fig. 1 of \cite{11}). The $\chi_{\omega}^{\prime} (T)$ response for a given $H$ is essentially dictated by $J_c (T)$. The temperature above which $\chi_{\omega}^{\prime} (T)$ (or, $J_c (T)$) becomes independent of the thermomagnetic history of the specimen appears to lie even above the (highest) peak temperature, i.e., $T_p^{zfc}$.

The pristine enunciation of the critical state model (CSM) prescribes a specific relationship between $J_c$ and the hysteretic magnetization response of a superconductor \cite{22}. It is well documented \cite{23, 24, 25, 26} that qualitative changes can occur in this relationship when $J_c(H)$ does not remain single valued function of H. For instance, the minor hysteresis loops display anomalous characteristics, like, the asymmetric shape, excursions beyond the envelope $M-H$ loop, etc. \cite{25, 26}, when the $J_c(H)$ turns path dependent. In the back drop of these observations, it is instructive to examine the response of the third harmonic of the ac susceptibility across the SMP and PE regions, i.e., from below the onset temperature of first of two anomalous variations in $J_c$ upto the irreversibility temperature ($T^{irr}$), where the (bulk) $J_c$ ceases. To be specific, consider the cool down of a weakly pinned type-II superconducting sample from above $T_c$ to below its $T^{irr}$, where the finiteness of $J_c$ would result in a non-linear magnetization response which could generate a measurable third harmonic signal in ac susceptibility measurements. Such a third harmonic signal would be expected to follow the increase in $J_c(T)$ for a given H, as per prescription of the CSM for path independent $J_c(H, T)$, as ($T^{irr}-T$) increases. The onset of the history dependence in $J_c(H)$ could compromise the above stated notion, arising from the applicability of the CSM. We continue to explore below, the limit of the path dependence in $J_c(H)$ via the observation of the history dependence in the first harmonic ac susceptibility data, and compare it with a specific feature noticeable in the temperature dependence of its third harmonic data.

Fig. 3(a) focuses attention onto the difference plots showing ($\chi_{\omega}^{\prime ZFC} - \chi_{\omega}^{\prime FCW}$) and ($\chi_{\omega}^{\prime ZFC} - \chi_{\omega}^{\prime FCC}$) at 10.5~kOe. The inset panel in Fig. 3(a) shows a portion of the data on an expanded scale to facilitate the marking of the limiting temperature $T^{\star}$, at which the larger of these differences vanish, and the $\chi_{\omega}^{\prime} (T)$ response becomes path independent. Fig. 3(b) presents the plot of $\chi_{3\omega}^{\prime ZFC} (T)$ and the difference plot, $\chi_{3\omega}^{\prime ZFC}(T) - \chi_{3\omega}^{\prime FCW}(T)$, in a field of 10.5~kOe. For the plots in Fig. 3 (b), we have identified the value of the limiting temperature $T^{\star}$ determined from Fig. 3 (a) and those of $T_{smp}^{on (zfc)}$, $T_p^{on (zfc)}$ and $T_p^{zfc}$ evident from Fig. 2 (c). Note that the $\chi_{3\omega}^{\prime ZFC} (T)$ data shows multiple undulations prior to reaching the limiting value $T^{\star}$. The first of these undulations appear to coincide with $T_{smp}^{on (zfc)}$, where the anomalous variation in $J_c$ commences. We recall that the association of an {\it enhancement} in the non-linear response with the occurence of anomalous variation in $J_c$ is well documented in the literature \cite{3}.

Next, we draw attention specifically to the behavior of $\chi_{3\omega}^{\prime}$ just above $T^{\star}$. Warming up from the low temperature side, as the temperature crosses the limit $T^{\star}$, $\chi_{3\omega}^{\prime} (T)$ is seen to monotonically decrease and vanish at the irreversibility temperature, $T^{irr}$ ($< T_c (H)$). There does not appear any simple correspondence between the $\chi_{\omega}^{\prime}$ and $\chi_{3\omega}^{\prime}$ for $T_{smp}^{on (zfc)} < T < T^{\star}$ (cf. Figs. 2(c) and 3(b)). While Fig. 2(c) shows that $\chi_{\omega}^{\prime} (T)$ monotonically decreases above $T_p^{zfc}$, reflecting the collapse in $J_c$ above the peak position of PE, $\chi_{3\omega}^{\prime} (T)$, on the other hand in Fig. 3 (b) appears to enhance between $T_p^{zfc}$ and $T^{\star}$. $\chi_{3\omega}^{\prime} (T)$ is seen to turn around only above $T^{\star}$ and follow the $J_c(T)$ thereafter (see Fig. 3(b)). As stated earlier, the $\chi_{3\omega}^{\prime} (T)$ signal can be related to the leading non-linear term in the prescription of CSM \cite{22}. It is therefore, {\it not fortuitous} that an imprint of the limit of the path independence in $J_c$ is present in the $\chi_{3\omega}^{\prime} (T)$ response.

To establish the assertion on the limit of the path independence in $J_c$ in the $\chi_{3\omega}^{\prime} (T)$ data, we have examined the above stated behavior at different fields and in different weakly pinned crystals of $2H$-$NbSe_2$ and found one to one correlation between $T^{\star}$ determined from $\chi_{\omega}^{\prime} (T)$ data in different thermomagnetic histories and the limiting temperature above which $\chi_{3\omega}^{\prime} (T)$ monotonically decreases. For instance, the Figs. 4(a) and 4(b) show $\chi_{3\omega}^{\prime} (T)$ data in fields of 3.6~kOe and 5~kOe in crystals $Z^{\prime}$ and $Z$ of $2H$-$NbSe_2$, respectively. The $T_p^{on}$, $T_p^{zfc}$ and $T^{\star}$ values (determined from difference ($\chi_{\omega}^{\prime ZFC} - \chi_{\omega}^{\prime FCC}$) plots) have been marked for each of the curves in Figs. 4(a) and 4(b). It is evident that $T^{\star}$ represents the limiting temperature above which $\chi_{3\omega}^{\prime} (T)$ monotonically decreases. At $T < T^{\star}$, the modulations in $\chi_{3\omega}^{\prime} (T)$ display complex behavior, dictated by $J_c(T)$ in different thermomagnetic histories and the $h_{ac}$ value in which the $\chi_{3\omega}^{\prime} (T)$ data are recorded. The $T^{\star}$, however, does not appear to vary in any noticeable manner with the amplitude of $h_{ac}$ (all data not shown here).

It is useful to explore the correlation between the $\chi_{3\omega}^{\prime} (T)$ and the noise signal in $\chi_{\omega}^{\prime} (T)$ \cite{16}, which can be easily recorded using a Lock-in amplifier (Stanford Research Inc. Model SR~850) having a flat band filter option \cite{27}. The above stated noise signal is believed to measure the fluctuations in $\chi_{\omega}^{\prime} (T)$ and it is argued \cite{16} to reflect the possibility of transformations amongst metastable states accessible from a given mode (ZFC/FC). Fig. 4(c) depicts the plot of noise in $\chi_{\omega}^{\prime ZFC} (T)$ in a field of 5~kOe and in $h_{ac} = 0.5~Oe$ (rms). We have marked the value of $T^{\star}$ alongwith the identification of corresponding values (in $h_{ac}$ of $0.5~Oe$ (rms) ) of $T_p^{on}$ and $T_p$ in Fig. 4(c). It is indeed not a coincidence that the noise signal recedes to the background value as $T \rightarrow T^{\star}$. Taking cue from earlier studies of noise in transport experiments \cite{28, 29}, Banerjee {\it et al} \cite{16} had argued that the increase in noise at $T = T_p^{on}$ ($\equiv T_{pl}$ in Ref. \cite{16}) reflects the possibility of enhancement in transformations amongst coexisting \cite{10} metastable states in a fractured (partially disordered) vortex solid. The setting in of the decrease in the noise signal as $T \rightarrow T_p$ was considered to imply the effect of phase cancellation of a large number of incoherent fluctuations as the vortex matter moves towards the fully disordered state. In such a framework, the possibility of coexistence of ordered pockets embedded in the disordered medium would cease as $T \rightarrow T^{\star}$, and the noise signal would reach the background value. In the context of our present results, the $T^{\star} (H)$ values represent the notion of spinodal line, $T_s (H)$ \cite{11}.

To summarize, we have collated together in Fig.5 the different (reduced) field-temperature (h, t) values for the crystal $Z'$. It includes in it the $t^{\star}$ ($= T^{\star}/T_c(0)$) values for the other crystals ($Y'$ and $Z$) of $2H$-$NbSe_2$ as well. For the sake of completeness, we have also chosen to depict in it the values of $h_{plat}(t)$ ($= H_{plateau}(T)/ H_{c2}^{\|c}(0)$, where $H_{c2}^{\|c}(0) = 44~kOe$) in the crystal $Z'$, as determined \cite{13, 14} from the normalized plots of the critical current density vs reduced fields (data not shown here). To recall \cite{13}, $H_{plateau}$ represents the limiting field below which the collective pinning regime gives way to the small bundle pinning regime. It seems appropriate to identify the (h, t) space between $h_{plat}(t)$ and the $t_{smp}^{on}(h)$ as the Bragg (elastic) glass region \cite{30}. Above the $t^{\star}(h)$ line, where the metastability effects cease, the vortex matter exists in the pinned amorphous phase.

The parameter space in between the elastic glass and the pinned amorphous state appears to belong to the co-existing weaker pinned (ordered) and stronger pinned (disordered) regions \cite{10}. At this juncture we draw attention again to the $\chi_{\omega}^{\prime}$ data in Figs. 2(c) (see also, $\Delta \chi_{\omega}^{\prime}$ data in Fig. 3(a)). These data reveal that as the temperature exceeds $T_{smp}^{on}$, the weaker pinned vortex solid in the ZFC warm-up mode starts to transform towards somewhat stronger pinned state (presumably with the injection and survival of disorder into the ordered state \cite{8, 9}) upto the peak temperature of the SMP anomaly. On the other hand, the transformation that sets in at $T_{smp}^{on}$ for the stronger pinned vortex solid state in the FCW mode, takes this state rapidly towards the weaker pinned (ordered) state upto the onset position of the PE. It is indeed only above $T_p^{on}$ that the shift towards the disordering commences for both, the ordered state in the ZFC mode and the disordered state in the FCW mode. While cooling down, the supercooling commences as temperature is lowered below $T^{\star}$, and it could appear to get further fortified as the temperature is lowered down across the SMP region (cf. Fig. 2(c)). {\it We believe that in the co-existing region, somewhere the balance shifts from the dominance of ordered regions to that of the disordered regions, in the sense that transformations triggered by rise in temperature and/or imposition of a large ac drive push a given state either towards a more ordered or, a more disordered state. {\bf We are tempted to suggest that the onset temperatures of the PE anomaly determine the said crossover boundary, which partitions the coexistence (h, t) space into region I ($t_{smp}^{on}~<~t<~t_p^{on}$) and region II ($t_p^{on}~<~t<~t^{\star}$)}} (see Fig. 5). In terms of the concept of the stationary state \cite{26}, $T_p^{on}$ represents the limiting temperature above which the disordered regions represent the stationary state.

Lastly, we attempt to make a contact with the theoretical results of Li and Rosenstein (LR) \cite{31} on the spinodal line, following the procedure of Ref. \cite{11}. In terms of the dimensionless scaled variables, LR theory gives the spinodal line as, $a_{LR}(t, h) = - [ (\pi^2 Gi) / 2 ]^{-1/3}[(1-t-h)t^{-2/3}h^{-2/3}]$, where $a_{LR}(t, h) = - 5$, and the rest of symbols have their usual meaning and representative values \cite{32, 33, 34, 35, 36}. For sample $Z^{\prime}$, if we take $H_{c2}^{\|c}(0) = 44~kOe$, $\epsilon=0.3$, $\xi = 8.6 nm$, $\lambda = 135 nm$ \cite{11, 36}, the theoretical spinodal line (solid line) satisfies our experimental data. The choice of $H_{c2}^{\|c}(0) = 52~kOe$ \cite{5} and $39.7~kOe$, for crystals $Y^{\prime}$ and $Z$, make the data points for these crystals also to satisfy the same spinodal line \cite{11}.

The vortex matter below the spinodal line in the phase diagram is typified by its inhomogenous nature of admixed weaker pinned (ordered) and stronger pinned (disordered) regions. The relative weights of these regions and their topographic distribution in different thermomagnetic histories elucidate the extent of measurable path dependence in $J_c(H, T)$. The presence of a driving force (ac field, transport current, electric field ($\frac{dH}{dt}$), etc.) can affect the relative weights of stronger and weaker pinned regions in a dynamical manner, and thereby influence the observed macroscopic magnetization response between the onset temperature of the anomalous variation in critical current density and the limiting $T^{\star}$ value. The peak temperature in anomalous variation in $J_c$ across the PE, is no longer viewed as the limiting temperature above which the history effects in $J_c$ cease \cite{3, 11, 16}. Xiao {\it et. al.} \cite{11} have stated that the peak temperature of the PE gets determined by the competition between the progressive nucleation of stronger pinned regions (which has the tendancy to enhance the sample averaged $J_c$ value) and the decrease in pinning strength with increase in temperature of both the stronger and the weaker pinned regions. To this, we may add that the non-linear responses that can independently arise from admixed stronger and weaker pinned regions undergo a qualitative change above $T^{\star}$, where only the homogenous stronger pinned regions (disordered phase) survive. The $T^{\star}$ value thus is expected to be independent of the amplitude of the ac driving force, whereas the peak temperature (and/or the history dependence in magnetization) value could depend on $h_{ac}$.

To conclude, our studies have shown that the PE is a very sharp transition in $2H$-$NbSe_2$ samples with very weak pinning, however, superheating/supercooling effects across PE are difficult to explore in such samples in presence of a driving force. In somewhat stronger pinned crystals of $2H$-$NbSe_2$, where the metastability effects manifest in a prominent manner, an imprint of the spinodal temperature can be conveniently located in the third harmonic data.

We have benefited from discussions with E. Y. Andrei, S. Bhattacharya, C. V. Tomy, A. Tulapurkar, D. Pal and D. Jaiswal-Nagar. We also thank P. L. Gammel for the crystal $Y^{\prime}$ and R. S. Sannbhadti and U. V. Vaidya for technical assistance. One of us (ADT) would like to acknowledge the TIFR Endowment Fund for the Kanwal Rekhi Career Development support. \\

$^{\star}$~~ajay@tifr.res.in, grover@tifr.res.in

\newpage
\begin{figure}
\includegraphics[scale=1.5,angle=0]{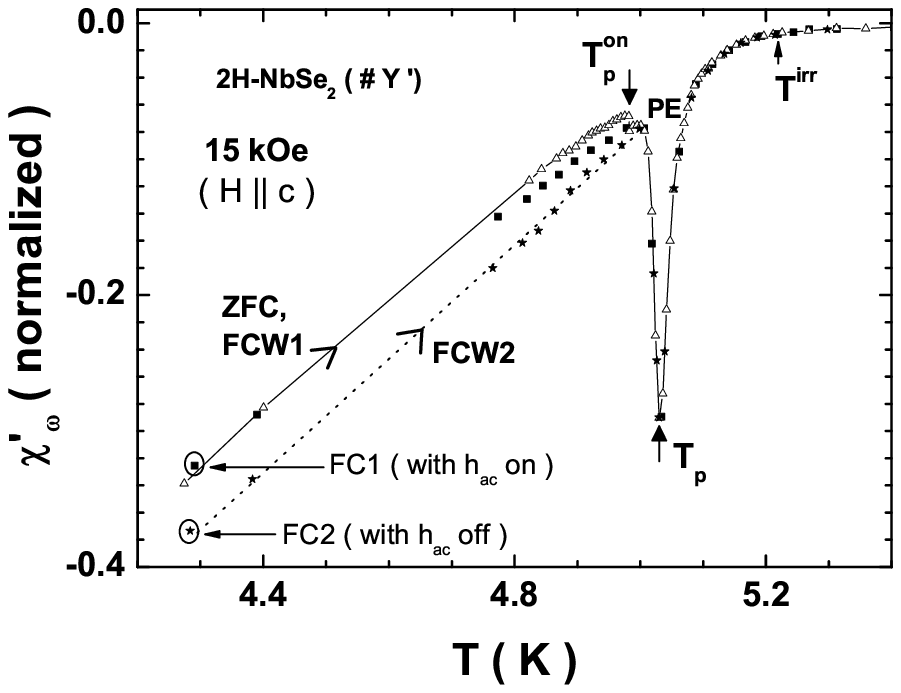}
\caption{In-phase ac susceptibility ($\chi_{\omega}^{\prime}$) data at $H = 15~kOe (\|c)$ and in $h_{ac}$ of 0.5~Oe (rms) in the crystal $Y^{\prime}$ ($T_c(0) \approx 7.25 K$) of $2H$-$NbSe_2$ for different thermomagnetic histories, as indicated. While obtaining the FC1 and FC2 states, the $h_{ac}$ remained {\it on} and {\it off}, respectively.}
\end{figure}
\begin{figure}
\includegraphics[scale=1.5,angle=0]{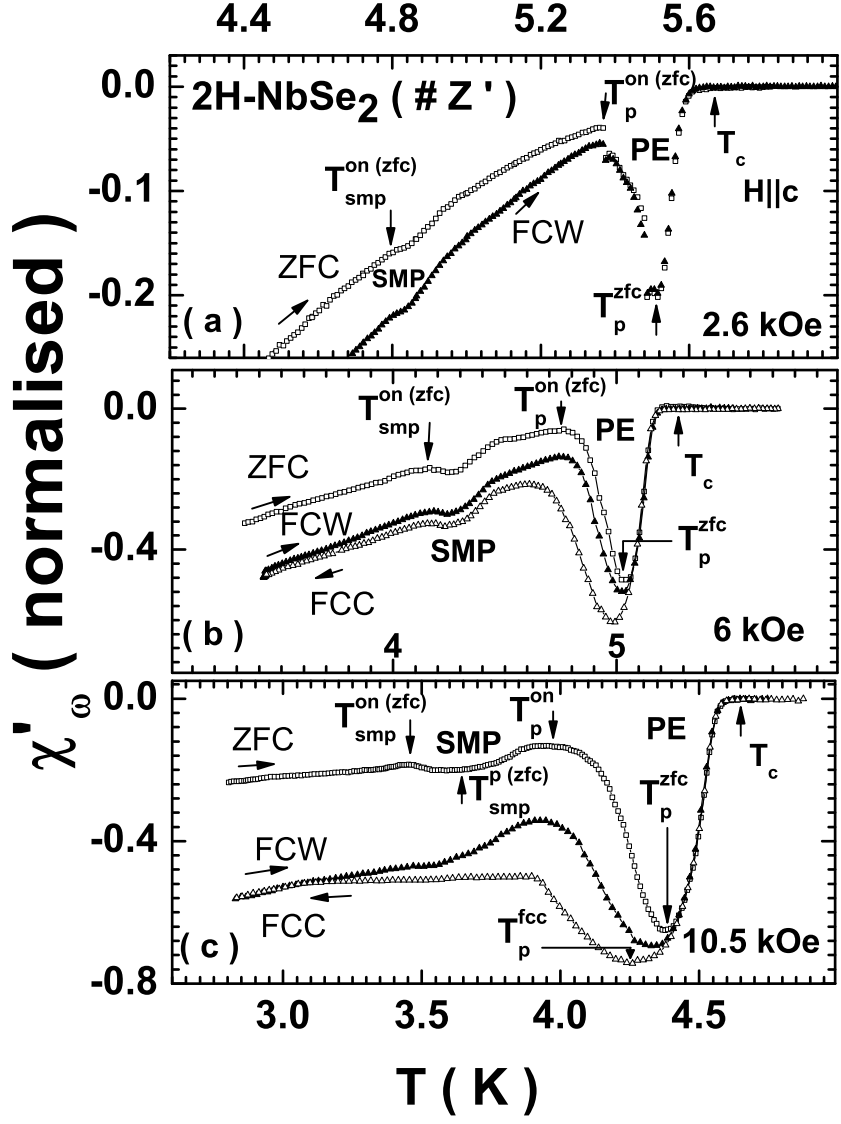}
\caption{$\chi_{\omega}^{\prime}(T)$ data ($h_{ac} = 1.5~Oe (rms)$) for different thermomagnetic histories at the dc fields ($H \| c$) indicated in the crystal $Z^{\prime}$ ($T_c(0) \approx 6.0 K$) of $2H$-$NbSe_2$. The anomalies corresponding to the SMP and the PE have been identified and the respective positions of $T_{smp}^{on}$, $T_p^{on}$, $T_p^{zfc}$ and $T_c$ have been marked.}
\end{figure}
\begin{figure}
\includegraphics[scale=1.5,angle=0]{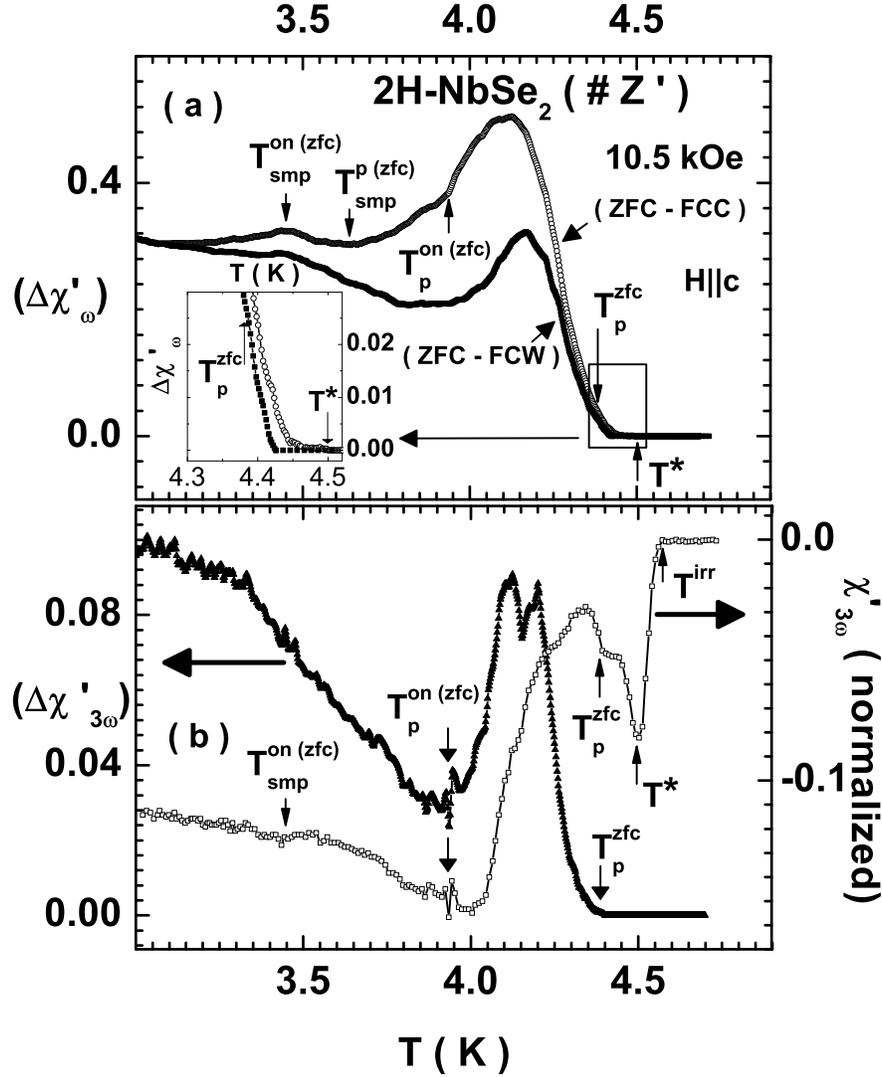}
\caption{The panel (a) depicts the difference plots ($\Delta\chi_{\omega}^{\prime}$) in the ZFC/FCC and ZFC/FCW modes at 10.5~kOe. The inset in panel (a) shows the data on an expanded scale to help locate the $T^{\star}$ value. The values of $T_{smp}^{on}$, $T_p^{on}$ and the $T_p^{zfc}$ identified from Fig. 2(c), have also been marked in the panel (a). The panel (b) shows the in-phase part of the ac susceptibility at $3\omega$, $\chi_{3\omega}^{\prime} (T)$, in a dc field of 10.5~kOe and $h_{ac}$ of 2.5~Oe (rms), in the ZFC mode. In addition, it also displays the difference plot, $\Delta \chi_{3\omega}^{\prime} (T) = \chi_{3\omega}^{\prime zfc} (T) - \chi_{3\omega}^{\prime fcw} (T)$. The value of $T^{\star}$ determined from Fig. 3 (a) and those of $T_{smp}^{on (zfc)}$, $T_p^{on (zfc)}$ and $T_p^{zfc}$ determined from Fig. 2(b), have been marked in the plots depicted in the panel (b).}
\end{figure}
\begin{figure}
\includegraphics[scale=1.5,angle=0]{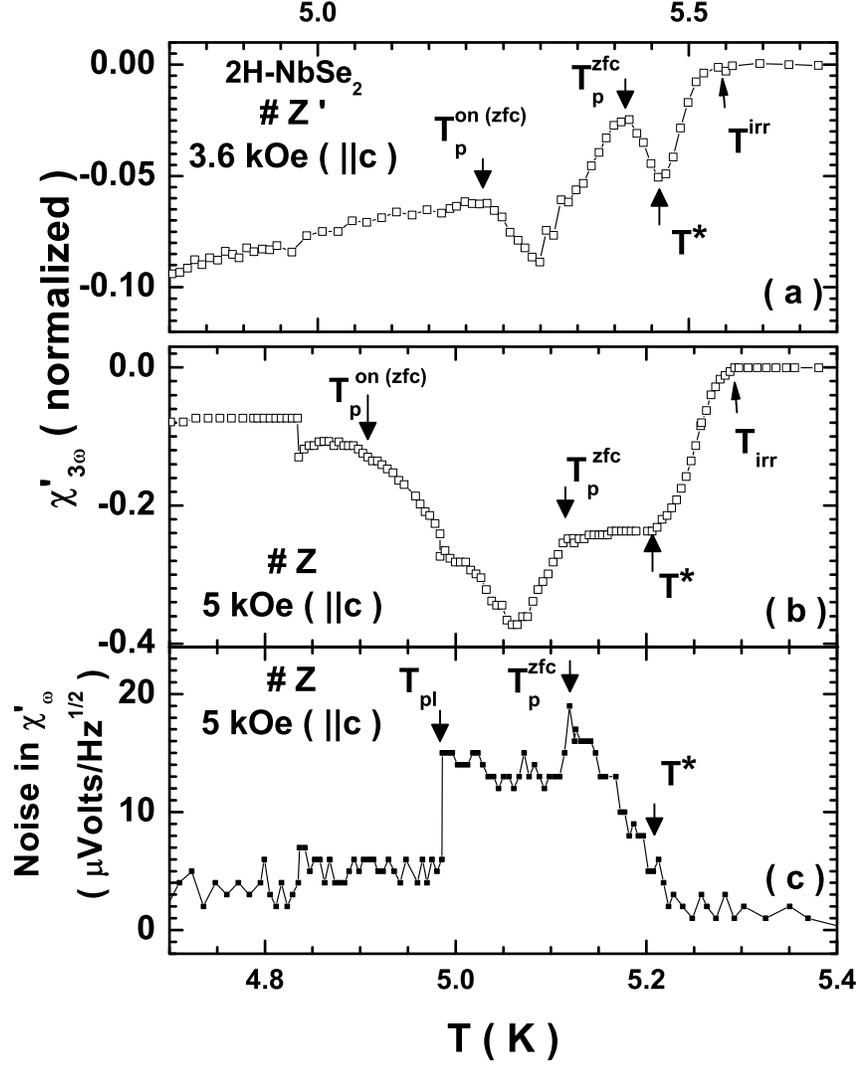}
\caption{The panels (a) and (b) show $\chi_{3\omega}^{\prime} (T)$ data at 3.6~kOe and 5~kOe for $h_{ac} = 2.5~Oe$ (rms) in the crystal $Z^{\prime}$ and $Z$, respectively. The panel (c) displays the noise in $\chi_{\omega}^{\prime} (T)$ recorded with a wide band filter at 5 kOe with $h_{ac} = 0.5~Oe (rms)$ and at a frequency of 211 Hz in the crystal $Z$. The positions of the respective $T_p^{on}$ ( \& $T_{pl}$ \cite{16}) and $T_p$ values (see text) have been marked in the different panels. }
\end{figure}
\begin{figure}
\includegraphics[scale=1.5,angle=0]{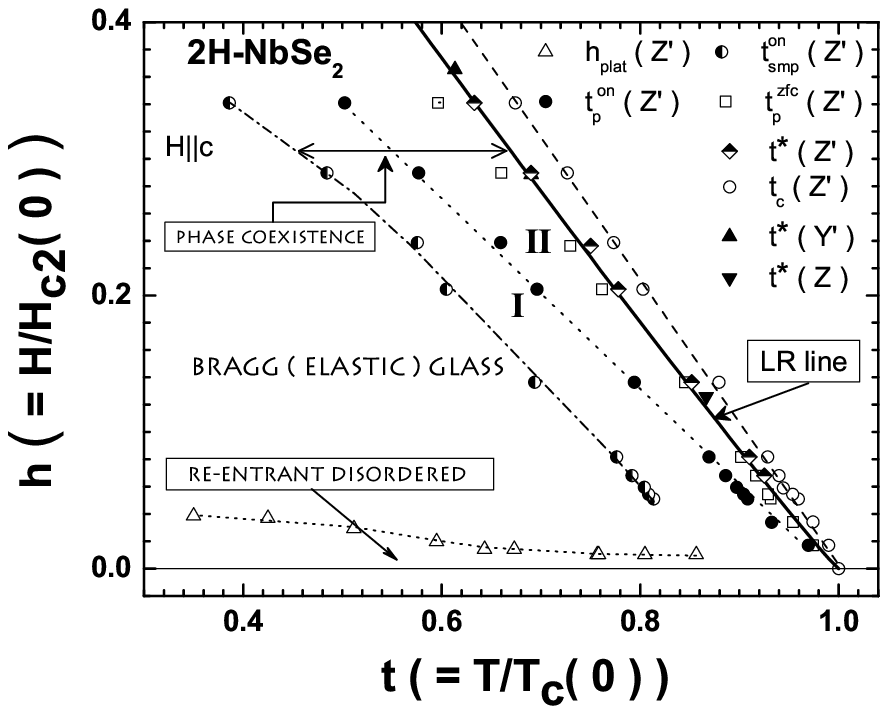}
\caption{A magnetic phase diagram in a typically weakly pinned single crystal $Z'$ of $2H$-$NbSe_2$ drawn in terms of reduced field ($h$) and temperature ($t$). The theoretical spinodal line based on the work of Li and Rosenstein \cite{31} has also been drawn as a solid line. The rest of the lines passing through various data sets are to guide the eye. Following Ref. \cite{11}, we have also included the data related to the limiting (spinodal) temperature in the other $2H$-$NbSe_2$ crystals ($Y^{\prime}$ and $Z$).}
\end{figure}

\begin{references}
\bibitem{27}
The noise in $Volts/ \surd Hz$ is simply the standard deviation divided by the square root of the equivalent noise bandwidth of the time constant. At each sample, the mean and moving average of the absolute value of the deviations is calculated. The averaging time (for the mean and moving average deviation) depends upon the time constant as is selected by the Lock-in amplifier (Model SR~850) in the range of 10 to 80 times the time constant.
\end{references}
\end{document}